\documentstyle[11pt,aaspp4]{article}

\def\ltsima{$\;\buildrel < \over \sim \;$}
\def\simlt{\lower.5ex \hbox{\ltsima}}
\def\gtsima{$\;\buildrel > \over \sim \;$}
\def\simgt{\lower.5ex \hbox{\gtsima}}
\begin{document}
\title{ISO observations of far-infrared rotational
emission lines of water vapor 
toward the supergiant star VY Canis Majoris $\bf ^*$}

\author{David A. Neufeld$^1$, Helmut Feuchtgruber$^2$, Martin Harwit$^3$, 
and Gary J. Melnick$^4$ }

\vskip 0.5 true in \parskip 6pt
  
\noindent{$^1$ Department of Physics \& Astronomy,  The Johns Hopkins University, 3400
North Charles Street,  Baltimore, MD 21218}
   
\noindent{$^2$ Max-Planck Institut f\"ur extraterrestrische Physik, 
Postfach 1603, D--85740, Garching, Germany}

\noindent{$^3$ 511 H Street S.W., Washington, DC 20024--2725; also Cornell University}

\noindent{$^4$ Harvard-Smithsonian Center for Astrophysics, 60 Garden Street, Cambridge, MA 02138}

\vskip 0.1 true in
{\parskip 6pt

\vskip 0.2 true in
\noindent{$^*$ Based on observations with ISO, an ESA project with instruments
     funded by ESA Member States (especially the PI countries: France,
     Germany, the Netherlands and the United Kingdom) with the
     participation of ISAS and NASA.}
\vskip 0.5 true in}

\keywords{circumstellar matter -- infrared: stars -- stars: abundances 
-- supergiants -- molecular processes -- stars: individual (VY Canis Majoris)}

\begin{abstract}

We report the detection of numerous far-infrared emission lines of
water vapor toward the supergiant star VY Canis Majoris.
A $29.5 - 45$ $\mu$m grating scan of VY CMa, obtained using the
Short Wavelength Spectrometer (SWS) of the Infrared Space Observatory 
(ISO) at a spectral resolving power $\lambda/\Delta \lambda$
of approximately 2000, reveals at least 
41 spectral features due to water
vapor that together radiate
a total luminosity $\sim 25 \,L_{\odot}$.  In addition to pure
rotational transitions within the ground vibrational state, these features
include rotational transitions within the (010) excited vibrational state. 
The spectrum also shows the $^2\Pi_{1/2}(J=5/2) \leftarrow ^2\Pi_{3/2}(J=3/2)$ 
OH feature near $34.6 \rm\,\mu m$ in absorption.  Additional SWS observations
of VY CMa were carried out in the instrument's Fabry-Perot mode 
for three water transitions:  the 7$_{25}-\,6_{16}\,$line at 
29.8367$\,\mu $m, the 4$_{41}-\,3_{12}\,$line at 31.7721\ $\mu $m, and the 
$4_{32}-\,3_{03}\,$ line at 40.6909 $\mu $m.  The higher spectral
resolving power $\lambda /\Delta \lambda $
of approximately $30,000$ thereby obtained 
permits the line profiles to be resolved spectrally for the first time
and reveals the ``P Cygni'' profiles that are characteristic of emission
from an outflowing envelope.

\end{abstract}

\section{Introduction}

A wealth of recent data from the Infrared Space Observatory 
(ISO; Kessler et al.\ 1996) has
demonstrated that water vapor is a ubiquitous constituent of 
the astrophysical universe.  With its powerful spectroscopic capability
unhindered by atmopheric absorption and covering 
the $2.38 - 197\,\rm \mu$m spectral region without interruption, ISO has 
detected water 
vapor in astrophysical environments as diverse as star-forming 
interstellar clouds (e.g. van Dishoeck \& Helmich 1996;
Helmich et al.\
1996; Ceccarelli et al.\ 1998; 
van Dishoeck et al.\ 1998); Herbig Haro ojects
(Liseau et al.\ 1996); shocked interstellar regions 
(e.g. Harwit et al.\ 1998); the
atmospheres of Jupiter, Saturn, Uranus, Neptune and Titan (Feuchtgruber
et al.\ 1997; Lellouch et al.\ 1997; Coustenis et al.\ 1998); 
the photospheres of cool stars 
(e.g. Tsuji et al.\ 1997); and the circumstellar
outflows from oxygen-rich evolved stars (e.g. Neufeld et al.\ 1996; 
Barlow et al.\ 1996, Justtanont et al.\ 1996).

The data now available from ISO observations support 
earlier theoretical predictions of a substantial abundance of water vapor 
within warm molecular astrophysical environments where the elemental 
abundance of oxygen exceeds that of carbon.  Since the water molecule 
possesses a large dipole moment, water rotational transitions play
an important role in the thermal balance of the gas 
(e.g. Neufeld, Lepp \& Melnick 1995).  

As part of a program to study water vapor in shocked interstellar
regions and circumstellar outflows, we have carried out a search
for far-infrared rotational emissions from water toward VY CMa, 
a high-mass red supergiant of spectral type M5Iae that is believed
to be in an advanced stage of evolution with a bolometric
luminosity $\sim 5 \times 10^5\,L_\odot$ and a mass loss rate that 
exceeds $10^{-4} M_\odot \rm \, yr^{-1}$ (Danchi et al.\ 1994).
A complete spectrum of VYCMa covering the
2.38 to 45 $\mu$m spectral region was obtained using the 
Short Wavelength Spectrometer (SWS; de Grauuw et al.\ 1996) on board ISO
and will be presented in a future publication; in this Letter we
consider only the 29.5 - 45 $\mu$m region (SWS Band 4) that contains
the strongest rotational emissions from water.  

\section{Observations}

All the observations of VY CMa reported here were carried out using the
SWS instrument.  A complete spectrum covering the 
2.38 to 45 $\mu$m spectral region was obtained using
the instrument's grating mode at a spectral resolving power
$\lambda / \Delta \lambda \sim 1000 - 2500$ (the 29.5 - 45 $\mu$m 
portion of which is presented here.)  In addition,   
the spectra of three rotational lines of ortho-water were obtained in
Fabry-Perot (FP) mode at a spectral resolving power
$\lambda / \Delta \lambda \sim 3 \times 10^4$:  
the 7$_{25}-\,6_{16}\,$ line at
29.8367$\,\mu $m, the 4$_{41}-\,3_{12}\,$ line at 31.7721$\,\mu $m, and the 
$4_{32}-\,3_{03}\,$ line at 40.6909$\, \mu $m. 

The observations were all carried out on 1997 November 19 (grating scan)
and 1997 November 20 (FP scans) with the ISO beam centered at position 
$\rm \alpha = 7^h 22^m 58.3^s, \delta = -25 ^\circ 46^\prime 03^{\prime\prime}$.
The beam size was $20 \times 33^{\prime \prime}$ for the grating scan
and $17 \times 40^{\prime \prime}$ for the FP scan.  Including overheads for
dark current measurements and calibration, the total observing times on target
were 10185 s for the complete 2.38 -- 45$\,\mu$m grating scan and 3962 s for
the three Fabry-Perot scans.

\section{Results}

The reduction of the grating spectrum was carried out using the
SWS Interactive Analysis system, applying calibration files
from version 7.0 of the ISO pipeline software.
The Fabry-Perot spectra were initially reduced using version 6.22 of the
ISO pipeline software, and the ISAP 
software package\footnote{The ISO Spectral Analysis Package (ISAP)
is a joint development by the LWS and SWS Instrument Teams and
Data Centers.  Contributing institutes are CESR, IAS, IPAC, MPE,
RAL and SRON.} was then used to remove bad data points and to 
co-add the individual spectral scans.

The $29.5 - 45 \rm \,\mu m$ grating scan of VY CMa, shown in
Figure 1, reveals a strikingly rich spectrum of water rotational 
transitions.  At least 41 features are readily identifiable as water,
corresponding to a density of more than 2 such features per
micron of wavelength coverage.  Line identifications and flux
estimates are given in Table 1.  Here we have been quite conservative
in identifying spectral features with water transitions, 
including only those features for which the identification seems most secure.

The flux calibration is believed
accurate to $\sim 30 \%$, and the typical 1$\sigma$ statistical
error in the flux is $\rm \sim several \times 10^{-20}\rm \, W \, cm^{-2}$.
Multiple scans of some portions of the spectrum 
strongly suggest that all
features apparent to the eye are real.  However, many features,
whether in absorption or emission, are still not reliably identified
(although many of them may, in fact, be water vapor lines).  As a 
result, the proper identification of a baseline, on top of which
individual features are observed, is quite uncertain.  For many
water lines, this difficulty substantially increases the 
uncertainties in the line flux estimates given in Table 1.  

In addition to transitions
within the ground vibrational state, the line list in Table 1
includes rotational transitions within the (010) excited
vibrational state.  Two radio wavelength transitions
within the (010) state had previously been detected
in VY CMa by Menten \& Melnick (1989).

The only molecule other than water for which
a feature has been identified in this spectral region is OH,
the primary photodissociation product of water, which we detected by means
of the $^2\Pi_{1/2}(J=5/2) \leftarrow ^2\Pi_{3/2}(J=3/2)$ 
absorption feature at $34.6 \,\rm \mu m$.

Several broad spectral features have been identified in the
20 -- 45 $\mu$m spectra of other evolved stars (e.g. Waters et al.\ 1996),
some of which have been tentatively attributed to crystalline silicates.
One broad spectral feature is prominent in the 29.5 -- 45 $\mu$m
VY CMa spectrum (Figure 1).  It is centered at wavelength 
$30.4 \pm 0.1 \mu$m, with a width $\sim 0.5\,\mu$m, and a
peak flux lying $\sim 3\%$ above the continuum. 
A feature with very similar properties has been detected in each of the sources
NML Cyg, AFGL 4106, HD 179821 and NGC 6302 (Waters et al.\ 1996): 
its identification is unknown at present.

Figure 2 shows the Fabry-Perot spectra obtained
toward VY CMa.  The corresponding line fluxes
are listed in Table 2: once again,
the likely errors in the flux calibration are believed to be
$\sim 30 \%$.

\section{Discussion}

\subsection{Grating spectrum}

The grating spectrum of VY CMa shown in Figure 1
demonstrates dramatically the ubiquitous nature of
water vapor emission lines in the far-infrared spectral
region.  Adopting 1500 pc
as the most credible estimate of the distance
to VY CMa (Lada \& Reid 1978), we find that the 
identified water features listed in Table 1 carry a total luminosity of
25 $L_{\odot}$.  Detailed modeling of the water line fluxes, which we defer to
a later publication, will be complicated significantly by large
departures from spherical symmetry.  In particular, recent Hubble
Space Telescope observations of VY CMa (Kastner \& Weintraub 1998)
vividly demonstrate the highly anisotropic nature of the surrounding
outflow.  These observations suggest that a highly flattened circumstellar
envelope obscures the photosphere of VY CMa from direct view along our 
line of sight, but that photospheric radiation does leak out along
other directions, giving rise to an elongated reflection nebula.
Clearly, this complex geometry is very different from the idealized
spherical geometry that has been assumed previously in modeling
water emissions from circumstellar outflows (e.g. Deguchi \& Rieu 1990,
Chen \& Neufeld 1995).  We expect that
future modeling efforts will be usefully constrained by additional
observations of the water vibrational bands near 2.7 and 6.2 $\mu$m
and of rotational absorption lines in the 16 -- 19.5 $\mu$m spectral region,
the reduction and analysis of which are underway. 

The strength of the 34.6 $\mu$m OH absorption feature is of
relevance to models for the radiative pumping of 1612 MHz 
OH masers by 34.6 $\mu$m infrared radiation (e.g. 
Elitzur, Goldreich \& Scoville
1976).  The OH absorption that we observed presumably arises
in a shell surrounding the source where the outflowing water
vapor is photodissociated by the interstellar radiation field.
Assuming spherical symmetry, we find
that the total rate of absorption in the OH 34.6 $\mu$m feature is
$\rm 1.0 \times 10^{47}$ photons per second (for an assumed
distance to the source of 1500 pc).  The average
maser photon emission rate is  
$\rm 4.3 \times 10^{45} \rm\,s^{-1}$ (Harvey et al.\ 1974),
with a typical time variability $\simlt 20 \%$.  Disregarding
the complications introduced by anisotropy, 
we therefore find that $\sim 23$ photons are
absorbed in the 34.6$\,\mu$m $^2\Pi_{1/2}(J=5/2) \leftarrow ^2\Pi_{3/2}(J=3/2)$
line for every 1612 MHz maser photon emitted.
The ratio of 34.6 $\mu$m photons absorbed to 1612 MHz photons
emitted is somewhat higher than
the value of 14 reported by Sylvester et al.\ (1997) for the evolved star
of IRC+10420.

\subsection{Fabry-Perot spectra of VY CMa}

The Fabry-Perot spectra of VY CMa shown in Figure 2
are of sufficiently high spectral resolution to yield useful 
information about the line profile.  As far as we are aware,
these are the first spectra in which the line
profiles of thermal water emissions from a 
circumstellar outflow have been resolved spectrally.
All three observed lines exhibit the classic ``P Cygni'' 
profiles associated with line formation in an outflow.
We find that the observed line profiles are consistent with
the line emission expected from a spherical outflow at
constant velocity, $v_{\rm out}$, accompanied by continuum
emission that suffers blueshifted absorption in the
approaching side of the outflow.   Figure 2 shows that
an outflow velocity of 
$v_{\rm out}=25 \rm \, km \, s^{-1}$ yields an acceptable fit
for all three lines: here the dashed line represents the symmetric,
parabolic emission line profile expected for optically-thick
line emission from a constant velocity outflow; the dotted line
represents continuum radiation with an optically-thick absorption
feature blueshifted by velocity $v_{out}$ relative to the
the emission line; and the solid line represents
the sum of those components after convolution with the
instrumental profile. 

The integrated fluxes in the emission and absorption
features, the equivalent widths of 
the absorption features, and the emission line
velocity relative to the LSR, $v_E$, are all tabulated in
Table 2.  The values of $v_{E}$ are in acceptable agreement
with each other and with the stellar LSR velocity of 
$\rm \sim 18 \,km\,s^{-1}$ inferred for this source
(Reid \& Dickenson 1976). 
The value of $25 \rm \, km\,s^{-1}$
obtained for $v_{\rm out}$ is close to -- but somewhat smaller
than -- the estimates of $\sim \rm 37 \, km \, s^{-1}$
and $\sim \rm 32 \, km\, s^{-1}$ inferred previously for the
terminal outflow velocity 
from observations of the thermal $J=2-1$ line of SiO
(Reid \& Dickenson 1976) and the 1612 MHz
OH maser line (Reid \& Muhleman 1978).  The observed spectra 
therefore provide an
important observational confirmation of theoretical predictions
(e.g. Deguchi \& Rieu 1990, Chen \& Neufeld 1995) 
that the far-infrared water line emission detected from
circumstellar outflows is generated largely
within a region where the outflow has already reached
-- or nearly reached -- its terminal velocity.
The absorption line equivalent width is considerably larger for
the $4_{32}-\,3_{03}\,$ line at 40.6909 $\mu$m than for
the other two lines, a difference that presumably results
from a large optical depth in this transition which has
a lower state of energy, $E_l/k$, only 197 K.

It is pleasure to acknowledge the support provided by
the Infrared Processing and Analysis Center (IPAC).
We also thank members of the help desk at Vilspa for their
assistance.  We gratefully acknowledge the support
of NASA grants NAG5-3316 to D.A.N.;
NAG5-3347 to M.H.; and NAG5-3542 and NASA contract NAS5-30702
to G.J.M.  \qquad  M.H.\ also thanks the Alexander von Humboldt foundation and
the Max Planck Institute for Radioastronomy for their hospitality
during a three-month visit to the MPIfR in Bonn.

\vfill\eject {\scriptsize 
\begin{tabular}{l c c c} 
\multicolumn{4}{c}{TABLE 1}\\
\multicolumn{4}{c}{Lines identified
in the grating scan of VY CMa}\\  \hline \hline \\
Line & Wavelength & Upper state & Flux  \\
     &            & Energy (E/k)&       \\
     &  ($\mu$m)  & (K)         & ($\rm 10^{-18}\,W\,cm^{-2}$) \\ 
\hline 
\\
H$_2$O  $ 7_{ 6 1} -  6_{ 5 2}$ & 30.526 & 1750 &  0.91\\
H$_2$O  $ 7_{ 6 2} -  6_{ 5 1}$ & 30.529 & 1750 & ---$^a$\\
H$_2$O  $ 4_{ 4 1} -  3_{ 1 2}$ & 31.772 & 702 &  0.65\\
H$_2$O  $ 6_{ 4 3} -  6_{ 1 6}$ & 32.313 & 1089 &  0.88\\
H$_2$O  $ 7_{ 5 2} -  6_{ 4 3}$ & 32.991 & 1525 & 2.22\\
H$_2$O  $ 6_{ 6 0} -  5_{ 5 1}$ & 33.005 & 1504 & ---$^a$\\
H$_2$O  $ 6_{ 6 1} -  5_{ 5 0}$ & 33.005 & 1504 & ---$^a$\\
H$_2$O  $10_{ 4 7} -  9_{ 3 6}$ & 33.510 & 2275 & 0.90\\
H$_2$O  $ 5_{ 5 0} -  5_{ 2 3}$ & 33.833 & 1068 & 0.82\\
H$_2$O  $ 9_{ 4 6} -  8_{ 3 5}$ & 34.396 & 1929 & 0.68\\
H$_2$O  $ 7_{ 3 4} -  6_{ 2 5}$ & 34.549 & 1212 & 0.58\\
OH $J=5/2- \leftarrow 3/2+$ $^b$ & 34.603 & 416 
& --1.49\phantom{--} \\
OH $J=5/2+ \leftarrow 3/2-$ $^b$ & 34.629 & 415 & --0.61\phantom{--} \\
H$_2$O  $ 7_{ 4 3} -  6_{ 3 4}$ & 35.429 & 1340 & 1.85\\
H$_2$O  $ 5_{ 3 3} -  4_{ 0 4}$ & 35.471 & 725 & 1.52\\
H$_2$O  $ 8_{ 4 5} -  7_{ 3 4}$ & 35.669 & 1615 & 0.94\\
H$_2$O  $ 6_{ 5 1} -  5_{ 4 2}$ & 35.904 & 1279 & 0.60\\
H$_2$O  $ 6_{ 5 2} -  5_{ 4 1}$ & 35.938 & 1279 & 1.21\\
H$_2$O  $ 7_{ 5 2} -  7_{ 2 5}$ & 36.046 & 1525 & 0.97\\
H$_2$O  $ \nu_2\,\, 5_{ 5 0} -  4_{ 4 1}$ & 36.161 & 3462 & 0.74\\
H$_2$O  $ \nu_2\,\, 5_{ 5 1} -  4_{ 4 0}$ & 36.163 & 3462 & ---$^a$\\
H$_2$O  $ 6_{ 2 4} -  5_{ 1 5}$ & 36.212 & 867 & 0.56\\
H$_2$O  $12_{ 3,10} - 11_{ 2 9}$ & 36.786 & 2824 & 0.40\\
H$_2$O  $ 7_{ 4 4} -  6_{ 3 3}$ & 37.566 & 1335 & 0.90\\
H$_2$O  $ 4_{ 4 1} -  4_{ 1 4}$ & 37.984 & 702 & 0.80\\
H$_2$O  $11_{ 3 9} - 10_{ 2 8}$ & 38.895 & 2439 & 0.20\\
H$_2$O  $ 7_{ 3 4} -  7_{ 0 7}$ & 39.045 & 1212 & 0.67\\
H$_2$O  $ 5_{ 5 0} -  4_{ 4 1}$ & 39.375 & 1068 & 2.31\\
H$_2$O  $ 5_{ 5 1} -  4_{ 4 0}$ & 39.380 & 1068 & ---$^a$\\
H$_2$O  $ 6_{ 4 2} -  5_{ 3 3}$ & 39.399 & 1090 & ---$^a$\\
H$_2$O  $12_{ 2,11} - 11_{1,10}$ & 40.016 & 2554 & 0.40\\
H$_2$O  $ 8_{ 4 4} -  8_{ 1 7}$ & 40.179 & 1628 & 0.77\\
H$_2$O  $13_{ 1,13} - 12_{ 0,12}$ & 40.188 & 2600 &  ---$^a$\\
H$_2$O  $13_{ 0,13} - 12_{ 1,12}$ & 40.189 & 2600 &  ---$^a$\\
H$_2$O  $ 6_{ 4 3} -  5_{ 3 2}$ & 40.337 & 1089 & 1.32\\
H$_2$O  $\nu_2\,\, 5_{ 4 1} -  4_{ 3 2}$ & 40.478 & 3240 & 0.90\\
H$_2$O  $\nu_2\,\, 5_{ 4 2} -  4_{ 3 1}$ & 40.687 & 3240 & 2.21\\
H$_2$O  $ 4_{ 3 2} -  3_{ 0 3}$ & 40.691 & 550 &  ---$^a$\\
H$_2$O  $ 6_{ 3 3} -  5_{ 2 4}$ & 40.760 & 952 & 1.17\\
H$_2$O  $11_{ 2 9} - 10_{ 3 8}$ & 40.894 & 2433 & 0.54\\
H$_2$O  $10_{ 3 8} -  9_{ 2 7}$ & 40.948 & 2081 & 0.63\\
H$_2$O  $13_{ 2,12} - 13_{ 1,13}$ & 42.427 & 2939 & 0.29\\
H$_2$O  $13_{ 1,12} - 13_{ 0,13}$ & 42.438 & 2938 &  ---$^a$ \\
H$_2$O  $ 9_{ 3 7} -  8_{ 2 6}$ & 42.859 & 1750 & 0.79\\
H$_2$O  $\nu_2\,\, 5_{ 2 3} -  4_{ 1 4}$ & 43.035 & 2955 & 0.65\\
H$_2$O  $11_{ 3 8} - 10_{ 4 7}$ & 43.125 & 2609 & 0.50\\
H$_2$O  $11_{1,10} - 10_{ 2 9}$ & 43.250 & 2194 & 0.78\\
H$_2$O  $12_{ 1,12} - 11_{ 0,11}$ & 43.339 & 2241 & 0.24\\
H$_2$O  $12_{ 0,12} - 11_{ 1,11}$ & 43.341 & 2241 &  ---$^a$ \\
H$_2$O  $\nu_2\,\, 7_{ 3 5} -  6_{ 2 4}$ ? & 43.713 & 3510 & 0.49\\
H$_2$O  $ 5_{ 4 1} -  4_{ 3 2}$ & 43.893 & 878 & 1.35\\
H$_2$O  $ 7_{ 4 3} -  7_{ 1 6}$ & 44.048 & 1340 & 1.15\\
H$_2$O  $ 5_{ 4 2} -  4_{ 3 1}$ & 44.195 & 878 & 1.14\\
H$_2$O  $ 8_{ 3 6} -  7_{ 2 5}$ & 44.702 & 1448 & 1.19\\
\hline 
\\
\multicolumn{4}{l}{$^a$ Line blended with previous feature.  Quoted
flux represents a sum over all blended lines.} \\
\multicolumn{4}{l}{$^b$ Cross-ladder $^2\Pi_{1/2} 
\leftarrow ^2\Pi_{3/2}$ transitions.}
\end{tabular}}

\vfill\eject

{\scriptsize
\begin{tabular}{l c c c c c c } 
\multicolumn{7}{c}{TABLE 2}\\
\multicolumn{7}{c}{Results of Fabry-Perot observations of VY CMa}
\\ \\ \hline \hline \\
Line & Rest & Upper state   &\multicolumn{2}{c}{Line Fluxes} & Absorption line    & $v_{E}\,\,^a$ \\

     & Wavelength $^b$  & Energy (E/k)  & Emission     & Absorption      & Equiv. Width & \\
     &       	  &               & $F_E$        & $F_A$           & $W_A$        & \\
     &  ($\mu$m)  & (K) &\multicolumn{2}{c}{($\rm 10^{-18}\,W\,cm^{-2}$) } & (km s$^{-1}$)& (km s$^{-1}$)  
\\
\\ \hline 
\\

H$_2$O  $7_{25} - 6_{16}$ & 29.8367 & 1126 & 2.35 & --1.94  & 11  & 22  \\
H$_2$O  $4_{41} - 3_{12}$ & 31.7721 & 702 & 2.41 & --1.71  & 11  & 20  \\
H$_2$O  $4_{32} - 3_{03}$ & 40.6909 & 550 & 3.69 & --2.20  & 32  & 18  \\

\\
\hline 
\\
\end{tabular}

\noindent $^a$ Central velocity of the emission feature, relative to the LSR

\noindent $^b$ From Toth (1991)

}

\begin{figure}
\plotfiddle{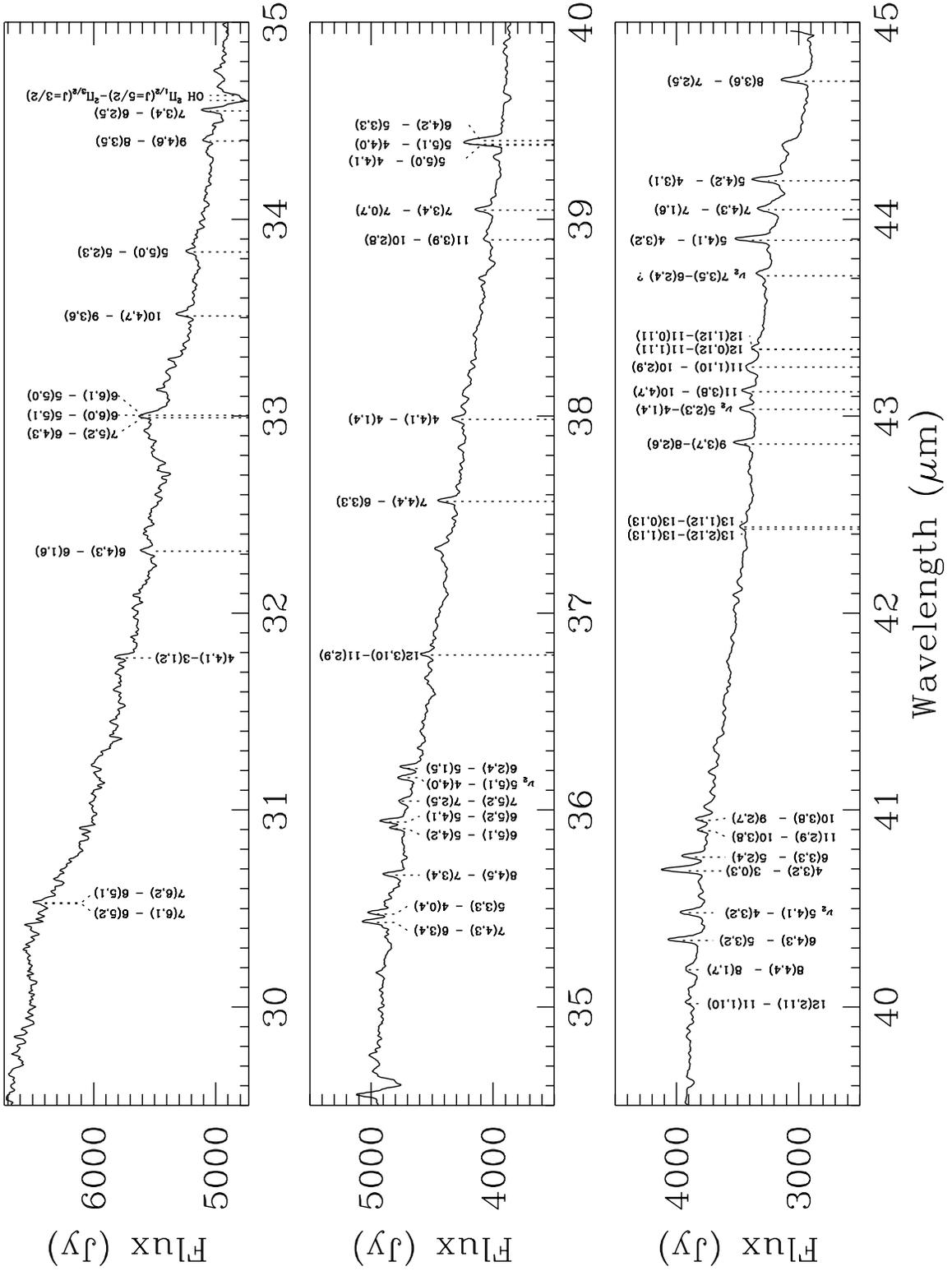}{570pt}{0}{80}{80}{-250}{-20}
\caption{29.5 -- 45$\,\mu$m SWS grating scan of VY CMa.  With the
exception of the OH absorption
feature at 34.6$\,\mu$m, all the identified features
are water lines.}
\end{figure}

\begin{figure}
\plotfiddle{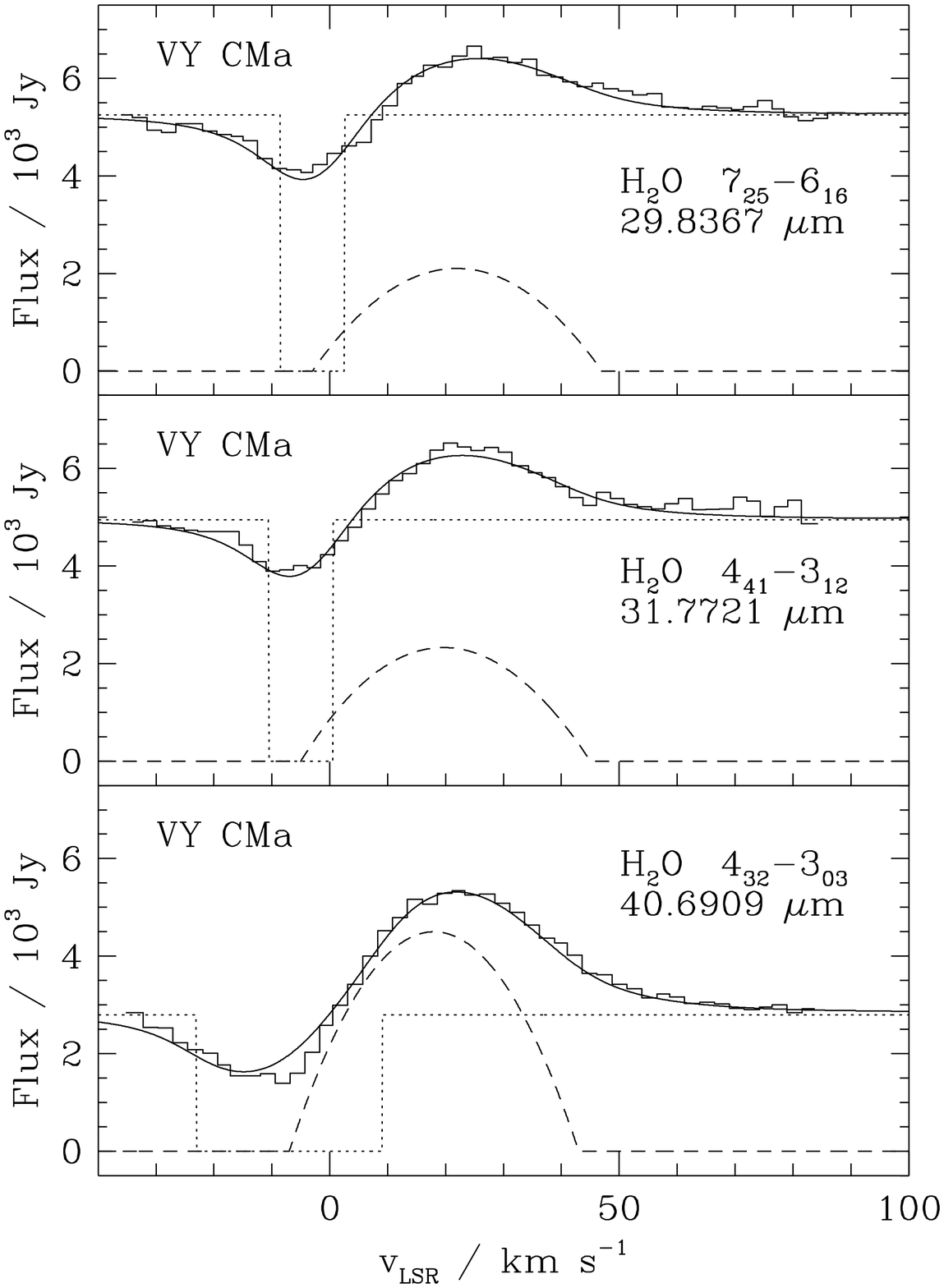}{550pt}{0}{80}{80}{-250}{-20}
\caption{ SWS Fabry-Perot spectra of three pure rotational lines of water
observed towards VY CMa:  
the 7$_{25}-\,6_{16}\,$ line at
29.8367$\,\mu $m, the 4$_{41}-\,3_{12}\,$ line at 31.7721$\,\mu $m, and the 
$4_{32}-\,3_{03}\,$ line at 40.6909$\, \mu$m.
Solid curves show the fits described in the text, obtained by
convolving the instrumental profile with the sum of the two
components represented by the dotted and dashed lines.}
\end{figure}


\begin{thebibliography}{}

\bibitem[]{} Barlow, M.J., et al. 1996, \aap, 315, L241 

\bibitem[]{} Ceccarelli, C., et al. 1998, \aap, 331, 372 

\bibitem[]{} Chen, W., \& Neufeld, D.A. 1995, ApJ, 452, L99 


\bibitem[]{} Clegg, P.E., et al. 1996, \aap, 315, L38 

\bibitem[]{} Coustenis, A., et al.\ 1998, A\& A, 336, L85

\bibitem[]{} de Graauw, T., et al. 1996, A\&A, 315, L49

\bibitem[]{} Danchi, W.C., Bester. M., Degiacomi, C.G., Greenhill,
L.J., \& Townes, C.H. 1994, AJ, 107, 1469

\bibitem[]{} Deguchi, S., \& Rieu, N. 1990, ApJ, 360, L27 

\bibitem[]{} Diamond, P.J., Norris, R.P., \& Booth, R.S. 1984,
MNRAS, 207, 611

\bibitem[]{} Elitzur, M., Goldreich, P., \& Scoville, N. 1976, ApJ, 205, 384

\bibitem[]{} Feuchtgruber, H., Lellouch, E., de Grauuw, T., Encrenaz, T., 
\& Griffin, M.\ 1997, Nature, 389, 159

\bibitem[]{} Goldreich, P., \& Scoville, N. 1976, ApJ, 205, 144 

\bibitem[]{} Harvey, P.M., Bechis, K.P., Wilson, W.J., 
\& Ball, J.S. 1974, ApJS, 27, 331

\bibitem[]{} Helmich, F.P., et al. 1996, \aap, 315, L173

\bibitem[]{} Justtanont, K., et al. 1996, \aap, 315, L217 

\bibitem[]{} Kastner, J.H., \& Weintraub, D.A. 1998, AJ, 115, 1592

\bibitem[]{} Kessler, M., et al. 1996, A\&A, 315, L27

\bibitem[]{} Harwit, M., Neufeld, D.A., Melnick, G.J., 
\& Kaufman, M.\ 1998, ApJ, 497, L105
 
\bibitem[]{} Lellouch, E., Feuchtgruber, H., De Graauw, T., 
Bezard, B., Encrenaz, T, \& Griffin, M. 1997, 
``First ISO Workshop on Analytical Spectroscopy", ESA, SP-419, p.\ 131

\bibitem[]{} Lada, C.J., \& Reid, M.J. 1978, ApJ, 219, 95

\bibitem[]{} Liseau, R., et al.\ 1996, A\&A, 315, L181

\bibitem[]{} Menten, K.M., \&  Melnick, G.J. 1989, ApJ, 341, L91

\bibitem[]{} Neufeld, D.A., Lepp, S., \& Melnick, G.J. 1995, ApJS, 100, 132

\bibitem[]{} Neufeld, D.A., et al. 
1996, \aap, 315, L237

\bibitem[]{} Reid, M.J., \& Dickenson, D.F. 1976, ApJ, 207, 784 

\bibitem[]{} Reid, M.J., \& Muhleman, D.O. 1978, ApJ, 220, 229

\bibitem[]{} Richards, A.M.S., Yates, J.A., \& Cohen, R.J. 1996, 
MNRAS, 282, 665

\bibitem[]{} Sylvester, R.J., et al. 1997, MNRAS, 291, L42 

\bibitem[]{} Toth, R. A. 1991, J Opt Soc Am B, 8, 2236

\bibitem[]{} Tsuji, T., Ohnaka, K., Aoki, W., \& Yamamura, I. 1997,
\aap, 320, L1 

\bibitem[]{} van Dishoeck, E.F. \& Helmich, F.P. 1996, \aap, 315, L177 

\bibitem[]{} van Dishoeck, E.F., et al.
1998, \apjl, 502, L173

\bibitem[]{} Waters, L.B.F.M., et al. 1996, \aap, 315, L361

\end{thebibliography}
\end{document}